\makeatletter\def\input@path{{short-paper/}}\makeatother
\documentclass[english]{lni}

\usepackage{graphicx}
\usepackage{latexml}
\usepackage{booktabs}
\usepackage{multirow}

\newcommand{\circled}[1]{%
  \iflatexml\textcircled{#1}\else
  \tikz[baseline=(char.base)]
    \node[shape=circle,fill=black,text=white,inner sep=1pt] (char) {\small #1};%
  \fi
}
\newcommand{\objectivecircle}[2]{%
  \textcolor{#1}{\iflatexml\textcircled{#2}\else\textcircled{\footnotesize #2}\fi}%
}

\definecolor{AstraOne}{HTML}{0072B2}
\definecolor{AstraTwo}{HTML}{D55E00}
\definecolor{AstraThree}{HTML}{009E73}
\definecolor{AstraFour}{HTML}{CC79A7}
\newcommand{\Aone}[1]{\textcolor{AstraOne}{\textbf{A1}}}
\newcommand{\Atwo}[1]{{\textcolor{AstraTwo}{\textbf{A2}}}}
\newcommand{\Athree}[1]{\textcolor{AstraThree}{\textbf{A3}}}
\newcommand{\Afour}[1]{\textcolor{AstraFour}{\textbf{A4}}}
\definecolor{GlmRun}{HTML}{6B5B00}
\newcommand{\Gone}[1]{\textcolor{GlmRun}{\textbf{G1}}}
\newcommand{\Gtwo}[1]{\textcolor{GlmRun}{\textbf{G2}}}
\newcommand{\Gthree}[1]{\textcolor{GlmRun}{\textbf{G3}}}
\newcommand{\Gfour}[1]{\textcolor{GlmRun}{\textbf{G4}}}

\newcommand{\dbtext}[0]{DBText}

\newcommand{\sparagraph}[1]{\vspace{1mm}\noindent {\bf #1}}

\usepackage{tikz}
\usetikzlibrary{backgrounds}
\usepackage{xcolor}

\usepackage{textcomp}
\usetikzlibrary{arrows.meta,calc}

\makeatletter

\tikzset{%
  fancy quotes/.style={
    text width=\fq@width pt,
    align=justify,
    inner sep=1em,
    anchor=north west,
    minimum width=\linewidth,
  },
  fancy quotes width/.initial={.5\linewidth},
  fancy quotes marks/.style={
    scale=5,
    text=white,
    inner sep=0pt,
  },
  fancy quotes opening/.style={
    fancy quotes marks,
  },
  fancy quotes closing/.style={
    fancy quotes marks,
  },
  fancy quotes background/.style={
    show background rectangle,
    inner frame xsep=0pt,
    background rectangle/.style={
      fill=gray!8,
      rounded corners,
    },
  }
}

\makeatother

\begin{document}

\title[The Case for Synthesized Data Compression]{The Case for Synthesized Data Compression}
 \author[1]{Ivan Iachnyk}{ivan.iachnyk@utn.de}{}
 \author[1]{Mihail Stoian}{mihail.stoian@utn.de}{}
 \author[1]{Luisa Neubauer}{luisa.neubauer@utn.de}{}
 \author[2]{Ryan Marcus}{rcmarcus@seas.upenn.edu}{}%
 \author[1]{Andreas Kipf}{andreas.kipf@utn.de}{}%
 \affil[1]{University of Technology Nuremberg, Germany}
 \affil[2]{University of Pennsylvania, USA}
\maketitle

\begin{abstract}

Compression is a fundamental tool in data management. Today's compression schemes are generally designed by hand to apply to many different types of data (e.g., run length encoding or arithmetic encoding), leading to a zoo of general-purpose algorithms which work well for many datasets, but may not be optimal for any specific dataset. Inspired by recent advancements in agentic coding, we propose synthesizing custom-tailored data compression algorithms on a per-dataset basis. Unlike prior work using agents to synthesize database components, a data-specific compression algorithm can be fully verified at construction time, alleviating most correctness concerns. Modern file formats, like AnyBlox and F3, allow including the synthesized code alongside the data itself. Experimentally, we show that our synthesis agent can expand the compression-factor/decompression-speed Pareto front, outperforming general-purpose methods by nearly $2\times$ in compression factor.
\end{abstract}


\section{Introduction}\label{sec:introduction}

\begin{figure}
\centering
\iflatexml
\includegraphics[width=\linewidth]{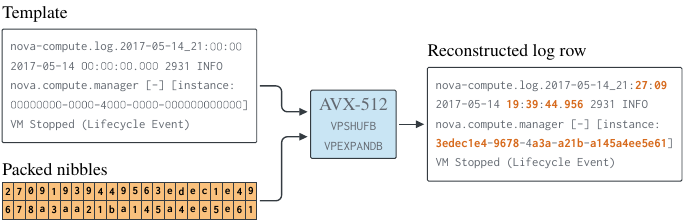}
\else
  \begingroup
  \resizebox{\linewidth}{!}{\input{short-paper/figures/openstack-run1-full-row.tikz}}%
  \endgroup

\fi
\caption{OpenStack \Aone{}: A fully synthesized compression codec for OpenStack log data. Constant characters are stored in templates, while changing timestamp and instance-ID digits are packed into four-bit values. Reconstruction uses AVX-512 nibble expansion and byte shuffling. The codec outperforms LZ4 in decompression speed (13 GB/s vs. 8 GB/s) and compression ratio (11$\times$ vs. 9$\times$).}
\label{fig:openstack-run1}
\end{figure}

Data compression is a cornerstone of efficient data management systems. In the past several decades, a myriad of general-purpose compression techniques have been designed and integrated into DBMSes~\cite{zstdRFC,fsst2020,exec_on_compressed}. The general strategy of our community has been to create a ``zoo'' of compression algorithms with sufficient diversity that any data encountered in the wild could be well served by at least one technique in the zoo. Here, we investigate an alternative route: what if we used LLM-powered agents to \emph{synthesize} custom compression codecs for specific pieces of data?

To illustrate why this may be desirable, consider the example in Figure~\ref{fig:openstack-run1}, showing a custom codec we synthesized for compressing OpenStack log data. The synthesizer noticed that most log messages are highly structured: they contain a timestamp and an instance ID. The synthesizer then designed a ``templated'' compression scheme that keeps constant characters in a template and densely packs the changing digits of timestamps and instance IDs into four-bit values. To make decoding efficient, our synthesizer designed an AVX-512 decompression scheme to reconstruct the original log entries.

This compression scheme is not particularly sophisticated or novel (in fact, similar variants have been previously described~\cite{clp2021}). The point is that the codec is custom-tailored to the underlying data. Our experiments show that it outperforms LZ4 in both decompression speed (13~GB/s vs. 8~GB/s) and compression ratio (11$\times$ vs. 9$\times$). A human expert could have built this codec without much trouble, but our system can generate it automatically.

\sparagraph{Correctness of Synthesized Systems.} The idea of databases custom-tailoring themselves to specific inputs is not new: instance-optimized systems, like learned indexing~\cite{kraska-li}, caching~\cite{schmidt-pc}, query optimization~\cite{kipf-lc, ryan1, ryan2, stoica-balsa}, and sorting~\cite{kraska-sorting}, represent this approach. But with the rise of coding agents, which generate code quickly and cheaply, our community has made good progress towards fully synthesizing a database system~\cite{wehrstein2026, gendb}. Unfortunately, guaranteeing correctness for such synthesized systems, especially across inputs that deviate from those seen by the synthesizer, remains challenging. 
One direction is formal methods, but this requires significant upfront effort in specification design and proof strategy, on which agents are happy to cheat. Another direction is rigorous testing, similar to the expansive test suites found in SQLite, but this can only increase our confidence in the synthesized code. It cannot demonstrate correctness outright.\footnote{One may argue that human-engineered systems also have no outright demonstration of correctness. While we believe the release and QA processes of most database vendors go far beyond testing, we note that this question is orthogonal to the present work.}

Here, we consider an alternative approach more aligned with the techniques used in instance-optimized systems: how can we build an interface such that high quality synthesized code can improve the performance of a system, but can never compromise the correctness of the system? Data compression offers a surprisingly simple solution: if we synthesize a codec for a specific block of data, we can execute the codec to ensure it correctly reconstructs the original data. If we only apply the synthesized codec to that exact data, we have now \emph{exhaustively} tested the codec, essentially proving its correctness.\footnote{Care must be taken to ensure the generated code is purely functional: dependencies on the system time, modifying state outside the codec's scope, or infinite loops could all still pose correctness issues. When instructed to avoid such properties, we found that agents complied 100\% of the time. Therefore, our model assumes a non-adversarial agent.} These synthesized codecs can then be integrated into modern file formats~\cite{vortex,anyblox} via code embedding (e.g.,~\texttt{WASM}).

\sparagraph{Prototype.} Our prototype codec synthesizer sits on top of an off-the-shelf coding agent and targets ever-present string data~\cite{amazon-tpc}. By guiding the agent with a simple experimental harness, we can target different trade-offs between compression ratio and decompression speed. We evaluate our prototype on five datasets, and find that our synthesized codecs extend the observed Pareto front: both the fastest and the most compact traditional baselines are each surpassed by a synthesized codec. For example, on the SQLStorm dataset, a synthesized codec exceeds bzip2-9\textquotesingle s~\cite{bzip2} compression factor by 49\% while decoding 34\% faster. These benefits also extend to individual database rows: on \dbtext{}, a synthesized codec achieves a 94\% higher compression factor than FSST~\cite{fsst2020} while processing selected rows 1.47--1.64$\times$ as fast. Our source code is publicly available at \href{https://github.com/utndatasystems/synthesized-data-compression}{\texttt{https://github.com/utndatasystems/}\allowbreak\texttt{synthesized-data-compression}}.

We present the architecture of our synthesizer in Section~\ref{sec:system}. In Section~\ref{sec:evaluation}, we compare synthesized codecs with state-of-the-art compression techniques. In Section~\ref{sec:analysis}, we examine a few of the generated codecs. Finally, we summarize related work in Section~\ref{sec:related-work}, and conclude in Section~\ref{sec:conclusion}.

\section{Synthesizing Compression Codecs}
\label{sec:system}

\begin{figure}
    \centering
    \includegraphics[width=0.7\linewidth]{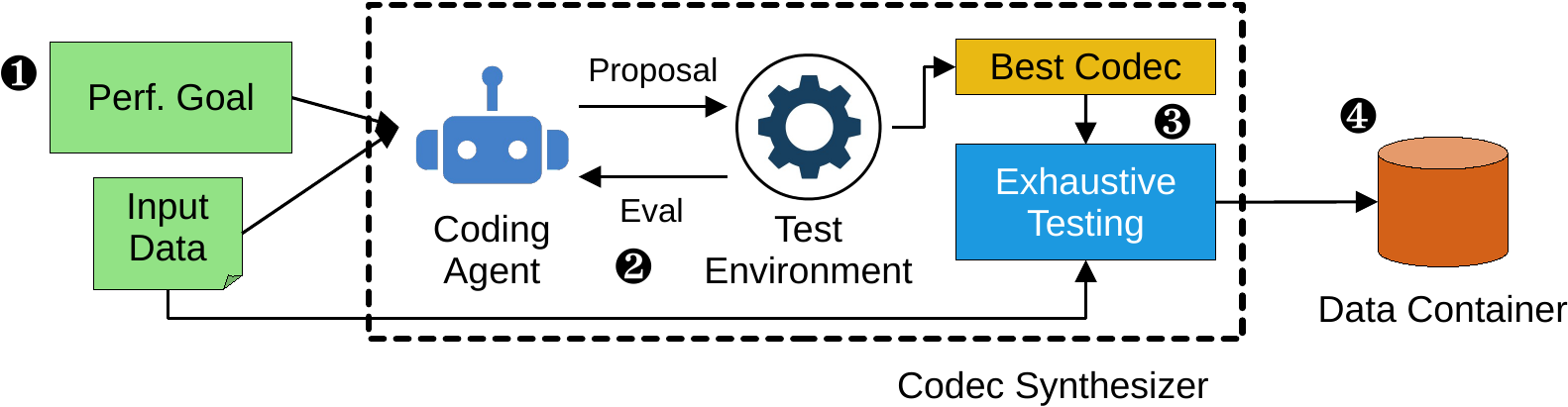}
    \caption{Codec synthesizer.}
    \label{fig:sysdiag}
\end{figure}

The architecture of our codec synthesizer is shown in Figure~\ref{fig:sysdiag}. First, a user specifies some input data to compress, along with a performance goal \circled{1}. Our current prototype is limited to textual input. A performance goal is an executable function encoding an objective. It could be absolute, like ``find the absolute best compression ratio possible,'' or relative, like ``beat ZSTD on decompression speed at a similar compression ratio.'' Then, we enter the synthesis loop \circled{2}. A coding agent is prompted to build a codec for the input data with the given performance goal. The agent can then submit a proposed codec to the test environment, which will measure the compression ratio, decompression speed, and correctness of the proposal. Eventually, when the agent declares victory or at a user-requested stopping point, the agent identifies the best-seen codec from prior proposals. That best codec is then exhaustively tested, ensuring that the decompressed data matches the input data byte-for-byte \circled{3}. If validation fails, the codec cannot be submitted, and the agent can revise it and repeat the evaluation. If the tests succeed, we can copy the proposed codec and compressed data into a data container (e.g., AnyBlox or Vortex) \circled{4}.

Our synthesis system is intentionally simple. Our goal is to leverage the intelligence of the coding agent and frontier models as much as possible. Next, we note two key design choices.

\sparagraph{Coding Agent Prompts.} Coding agents are notoriously sensitive to prompts. Our exact prompts are long, and are released with this paper's artifacts. Here, we highlight two key lessons learned while designing these prompts. First, it is important to tell the coding agent exactly what it should and should not do. Without clear instructions, we found that coding agents often tried to design new general-purpose compression algorithms (a significantly harder task). Instead, we needed to instruct the agent to design a codec \emph{only for these input data}, with an explicit instruction to overfit. Second, we found it very beneficial to provide reference values for compression factor and decompression throughput, as they gave the agent concrete targets to beat and encouraged further improvements. However, we did not name the algorithms behind these values, so the agent would remain free to explore different strategies instead of focusing on fine-tuning a particular baseline.

\sparagraph{Data Containers.} Modern data containers like Vortex~\cite{vortex}, AnyBlox~\cite{anyblox}, and F3~\cite{f3} support embedding decompression codecs directly inside of a data container. This feature was originally motivated by backwards compatibility, but we make use of it here to instead store our synthesized codecs. This means that storing data compressed with the codec requires storing both the compressed data and the codec itself, so the agent must simultaneously minimize both (a sort of Kolmogorov complexity~\cite{kolmogorov}). While we simply let the agent balance decompressor complexity and compression ratio, investigating a compact domain specific language for specifying decompressors is an interesting direction for future work.

\section{Experiments}\label{sec:evaluation}

We structure our experimental evaluation to answer a few key questions: (\textbf{RQ1}) Can a synthesizer generate competitive codecs in terms of compression ratio and decompression speed? (\textbf{RQ2}) Can a synthesizer produce codecs with competitive random access (i.e., indexing) speeds? (\textbf{RQ3}) Do factors like model choice, providing performance baselines, and allowing the use of external dependencies help the synthesizer?

\sparagraph{Setup.} Following prior work, we test compression on several datasets: OpenStack system logs (62~MB), PyTorrent~\cite{pytorrent} code (100~MB), Yelp~\cite{yelp} (100~MB), SQLStorm's~\cite{sqlstorm2025} TPC-DS queries (19~MB), and \dbtext{} (23 columns from FSST~\cite{fsst2020}, 39.84~MB). We run 4 synthesis attempts for each dataset focusing on a different performance tradeoff: \objectivecircle{AstraOne}{1}~Maximize decompression speed, \objectivecircle{AstraTwo}{2}~maximize compression factor, \objectivecircle{AstraThree}{3}~medium compression factor, maximize decompression speed, and \objectivecircle{AstraFour}{4}~medium decompression speed, maximize compression factor.

We compare against the following codecs: Zstandard~\cite{zstdRFC}, zlib~\cite{zlibRFC}, XZ~\cite{xzUtils}, LZ4~\cite{lz4format}, Brotli~\cite{brotliRFC}, bzip2~\cite{bzip2}, FSST~\cite{fsst2020}, and OnPair+~\cite{onpair2025,tokenTables2026}. For coding agents, we use (1) GPT-6 Astra via Codex CLI at \emph{Ultra} and (2) GLM-5.3-Flash via ZCode at \emph{Max}. A shared MCP interface supports inspection, implementation, benchmarking, and byte-exact checks for iterative revision. \textbf{Standalone} codecs use no external compression libraries, while \textbf{external-dependency} runs may use them. We test on an Intel Xeon Gold 5318Y (2.10~GHz) machine with 128~GB RAM, running Ubuntu 24.04.1 LTS. We measure compression factor as the original uncompressed data size divided by the compressed data size (for synthesized codecs, this includes compressed data and the decompression program; for traditional codecs, this is only the compressed data size), dictionaries, and indexes.

\begingroup
\setlength{\intextsep}{4pt}
\setlength{\floatsep}{6pt}
\setlength{\textfloatsep}{6pt}

\subsection{(RQ1) Can We Synthesize Competitive Block-Based Compression Codecs?}\label{subsec:block-based-results}

\begin{figure}[!tb]
    \centering
    \includegraphics[width=0.95\linewidth]{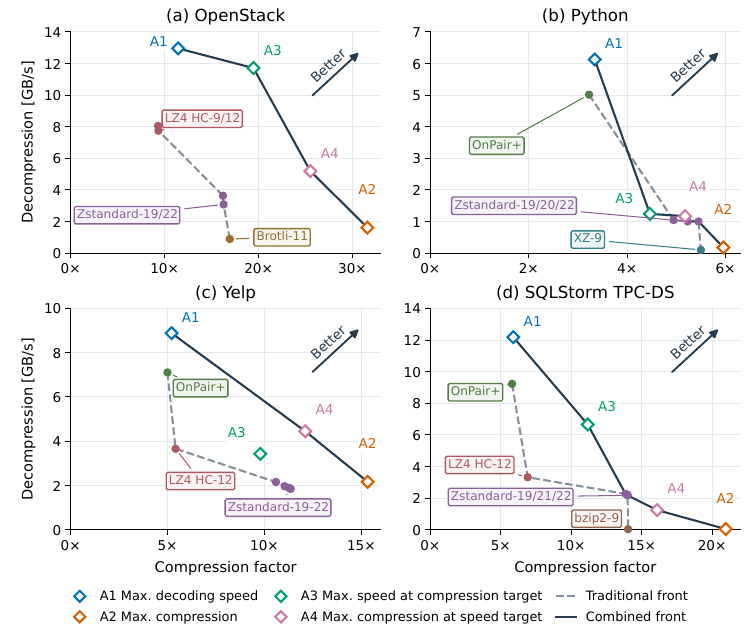}
    \caption{Block-decompression Pareto fronts of traditional compressors and standalone Astra codecs A1--A4, corresponding to the four objectives. Only traditional compressors on their front are shown.}
      \label{fig:block-comparison}
\end{figure}

We show the results on the four structured datasets in Figure~\ref{fig:block-comparison}.\footnote{For these datasets, Astra codecs compress at 0.9--769~MB/s, compared with 0.5--1,262~MB/s for the general-purpose baselines.} The synthesized codecs extend the observed Pareto front: on every dataset, we find codecs that match the best traditional compression factor or decompression speed while improving on the other. \Aone{}, which represents our synthesis run seeking to maximize decompression speed while at least matching the compression factor of the fastest traditional baseline, decodes 22--61\% faster than the fastest traditional baseline on each dataset while also achieving higher compression factors. At the other end, \Atwo{} exceeds the most compact baseline's compression factor by 8--87\% and still decodes faster.

These gains also extend to \dbtext{}, where each codec handles all 23 heterogeneous columns (Fig.~\ref{fig:dbtext-block-based}). Standalone \Aone{} decodes about 30\% faster than OnPair+ with slightly better compression. Meanwhile, \Atwo{} improves bzip2-9's compression factor by 7\% while decoding 8.4$\times$ as fast. Zstandard at high compression levels remains competitive between these endpoints.

\begin{figure}
  \centering
  \includegraphics[width=0.92\linewidth]{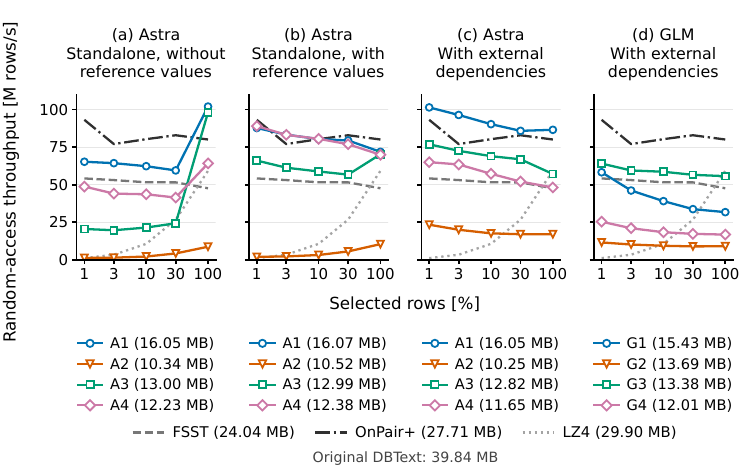}
  \caption{DBText random access: standalone Astra without (a) and with (b) reference values, and Astra (c) and GLM (d) with external dependencies and reference values.}
  \label{fig:dbtext-random-access}
\end{figure}

\subsection{(RQ2) Can We Synthesize Codecs with Fast Random Access?}\label{subsec:random-access-results}

Next, we switch from synthesizing codecs that decompress blocks of data at a time to synthesizing codecs that support random access. For these experiments, exhaustive testing means testing that random access into every possible position returns the correct result. We compare our random access codecs with FSST~\cite{fsst2020} and OnPair+~\cite{onpair2025} (adapted for random access with added row offsets), two state-of-the-art approaches, on the \dbtext{} data. We vary selectivity from 1\% to 100\% of rows and report the median of geometric-mean throughput across the 23 columns. 

In Fig.~\ref{fig:dbtext-random-access}(c), \Aone{} processes selected rows 1.67--1.87$\times$ as fast as FSST and also outperforms OnPair+ at every tested selectivity, while occupying only 16.05~MB, compared with 24.04~MB for FSST and 27.71~MB for OnPair+. At the same time, \Afour{} more than halves FSST's footprint while matching or exceeding its throughput. However, pushing compression further comes at a cost: \Atwo{} produces the smallest footprint, at 10.25~MB, but processes only 17--23 M rows/s.

\subsection{(RQ3) Ablation Studies}\label{sec:dbtext-ablations}

The standalone configurations in Figure~\ref{fig:dbtext-block-based} show the importance of providing \textbf{reference values} to the synthesizer. Without a clear target, the synthesizer does not know when to declare victory, or give up on one approach and try another.

The standalone and external-dependency configurations in Figure~\ref{fig:dbtext-block-based} show the importance of providing access to well-tuned versions of \textbf{external dependencies}, both for making comparisons and for remixing different ideas from prior work.  For example, the most compact synthesized codec combines front coding~\cite{stringDictionaries2016}, PPMd, and Brotli. Without external dependencies, Astra can already beat FSST in both compression factor and decompression speed, but allowing them helps it find more compact representations.

\begin{figure}
  \centering
  \includegraphics[width=\linewidth]{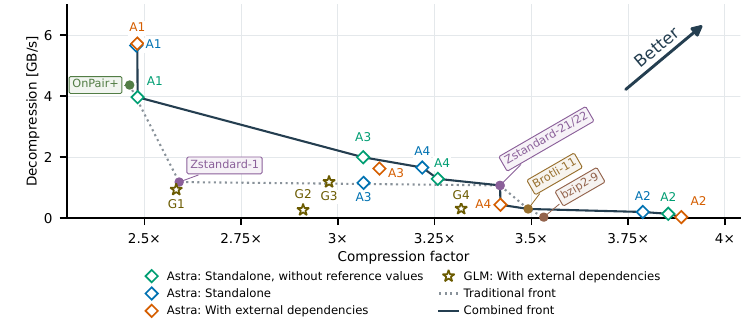}
  \caption{Synthesized codecs by GPT-6 Astra and GLM-5.3-Flash on \dbtext{}.}
  \label{fig:dbtext-block-based}
\end{figure}

While most of our experiments use Astra, we also tested the GLM model. In our setup, \textbf{standalone} GLM attempts did not produce byte-exact codecs, even after multiple continuation prompts. This is evidence that model quality is essential for synthesis. With external dependencies and reference values, however, GLM produced four byte-exact codecs, although only \Gthree{} met its performance targets. While \Gthree{} achieves a higher compression factor and faster random access than FSST (Fig.~\ref{fig:dbtext-block-based} and Fig.~\ref{fig:dbtext-random-access}(d)), every evaluated GLM codec is surpassed by an Astra codec in both compression factor and decompression speed.

\sparagraph{Cost.} At September 2026 API prices, estimated API-equivalent costs for 28 Astra runs range from \$20.03 to \$98.50 (median \$43.15), while four GLM runs range from \$5.58 to \$15.85. Decoding requires no model calls, allowing repeated uses to amortize synthesis. We expect model cost to drop over time (in fact, model costs dropped twice while we were conducting our experiments -- these calculations reflect that).

\endgroup

\section{Analysis of Synthesized Code}\label{sec:analysis}

In this section, we explore some of the most interesting ideas the agents came up with and show how they lead to better compression and faster decoding.

\subsection{System Logs}
\label{sec:results-openstack}

OpenStack log lines look different at first glance, but much of their content repeats. In \Aone{}, Astra groups lines into templates by masking numeric fields and hexadecimal identifiers~\cite{clp2021}. It then checks which characters actually change within each group. Even identifiers that differ between lines can share constant characters, which remain in the template. Each line therefore stores only a template identifier and its varying characters, packed into four-bit values. The decoder converts these values back to ASCII and inserts them into the template using AVX-512 masked expansion (Fig.~\ref{fig:openstack-run1}).

\Athree{} takes this idea a step further by generating dedicated decoding functions for common field layouts. Each function has the field positions and widths built in, allowing it to copy fixed text and restore timestamps, numeric values, and UUIDs directly. At the most compact endpoint, \Atwo{} achieves an 87\% higher compression factor than Brotli-11 while decoding approximately 1.8$\times$ as fast (Fig.~\ref{fig:block-comparison}(a)).

\subsection{Python Code}
\label{sec:results-python}

Python code offers another opportunity for a custom compressor, since its indentation often follows predictable patterns. A line ending in a colon, for example, is often followed by a line indented four spaces further. In \Atwo{}, Astra takes advantage of this by separating the leading spaces from the rest of each line and recording how the indentation changes. To compress these changes, it draws on three clues from the code: how the previous line ends, the current statement type, and the previous indentation depth. These clues divide the changes into 729 contexts, within which they are compressed using rANS~\cite{duda2013}.

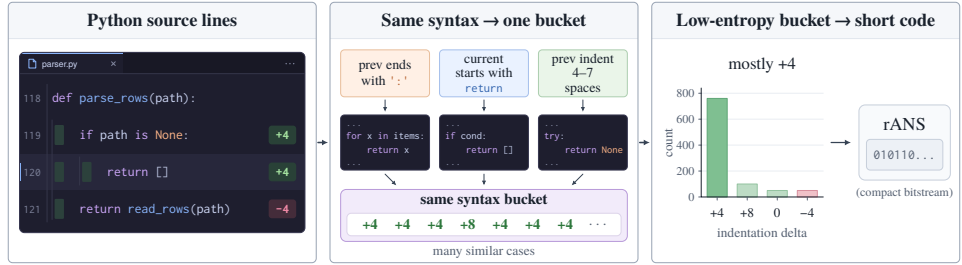
\begin{figure}[!htbp]
  \centering
  \begingroup
  \resizebox{\linewidth}{!}{\begingroup%

\definecolor{bcInk}{HTML}{19212C}%
\definecolor{bcMuted}{HTML}{566172}%
\definecolor{bcBorder}{HTML}{98A3B3}%
\definecolor{bcHeader}{HTML}{F5F7FA}%
\definecolor{bcRule}{HTML}{E1E6ED}%
\definecolor{bcArrow}{HTML}{364354}%
\definecolor{bcGrid}{HTML}{E9EDF2}%
\definecolor{bcPeach}{HTML}{FFF0E5}%
\definecolor{bcPeachEdge}{HTML}{E8BE9B}%
\definecolor{bcPeachInk}{HTML}{B96837}%
\definecolor{bcBlue}{HTML}{EAF2FF}%
\definecolor{bcBlueEdge}{HTML}{A5BFEB}%
\definecolor{bcBlueInk}{HTML}{3264AE}%
\definecolor{bcMint}{HTML}{EAF5E9}%
\definecolor{bcMintEdge}{HTML}{ACCEAF}%
\definecolor{bcLavender}{HTML}{F2EBFA}%
\definecolor{bcLavenderEdge}{HTML}{B89AD9}%
\definecolor{bcWellEdge}{HTML}{DDD4EA}%
\definecolor{bcGreenInk}{HTML}{2D773C}%
\definecolor{bcEditor}{HTML}{1E1E2E}%
\definecolor{bcEditorTop}{HTML}{181825}%
\definecolor{bcEditorEdge}{HTML}{36364F}%
\definecolor{bcEditorRule}{HTML}{35364A}%
\definecolor{bcActiveRow}{HTML}{27273B}%
\definecolor{bcCode}{HTML}{CDD6F4}%
\definecolor{bcCodeMuted}{HTML}{7F849C}%
\definecolor{bcPunctuation}{HTML}{BAC2DE}%
\definecolor{bcKeyword}{HTML}{CBA6F7}%
\definecolor{bcFunction}{HTML}{89B4FA}%
\definecolor{bcConstant}{HTML}{FAB387}%
\definecolor{bcPositive}{HTML}{A6E3A1}%
\definecolor{bcNegative}{HTML}{F38BA8}%
\definecolor{bcIndent}{HTML}{2C413C}%
\definecolor{bcPositiveWell}{HTML}{294137}%
\definecolor{bcNegativeWell}{HTML}{442C3A}%
\definecolor{bcBarStrong}{HTML}{7FBF91}%
\definecolor{bcBarStrongEdge}{HTML}{5EA676}%
\definecolor{bcBarSoft}{HTML}{BDDCC5}%
\definecolor{bcBarSoftEdge}{HTML}{94C3A1}%
\definecolor{bcBarRed}{HTML}{F0C1C8}%
\definecolor{bcBarRedEdge}{HTML}{D78F9D}%
\definecolor{bcEncoder}{HTML}{F8FAFC}%
\definecolor{bcEncoderEdge}{HTML}{C7CFDB}%
\definecolor{bcBitWell}{HTML}{EDF1F6}%

\newcommand{\BCserif}[1]{\fontencoding{T1}\fontfamily{qtm}\fontseries{m}\fontshape{n}\fontsize{#1}{#1}\selectfont}%
\newcommand{\BCsans}[1]{\fontencoding{T1}\fontfamily{qhv}\fontseries{m}\fontshape{n}\fontsize{#1}{#1}\selectfont}%
\newcommand{\BCmono}[1]{\fontencoding{T1}\fontfamily{zi4}\fontseries{m}\fontshape{n}\fontsize{#1}{#1}\selectfont\spaceskip=.5em\xspaceskip=.5em}%
\newcommand{\BCkw}[1]{\textcolor{bcKeyword}{#1}}%
\newcommand{\BCfn}[1]{\textcolor{bcFunction}{#1}}%
\newcommand{\BCvar}[1]{\textcolor{bcCode}{#1}}%
\newcommand{\BCconst}[1]{\textcolor{bcConstant}{#1}}%
\newcommand{\BCpn}[1]{\textcolor{bcPunctuation}{#1}}%
\newcommand{\BCcode}[4]{\node[anchor=base west,font=\BCmono{#3},text=bcCode] at (#1,#2) {#4};}%
\newcommand{\BClabel}[4]{\node[anchor=base,font=\BCserif{#3}] at (#1,#2) {#4};}%

\begin{tikzpicture}[
  x=1pt,y=-1pt,
  every node/.style={inner sep=0pt,outer sep=0pt,text=bcInk},
  line cap=round,line join=round,
  bc flow/.style={draw=bcArrow,line width=2.5pt,-{Latex[length=10pt,width=9.5pt]}},
  bc branch/.style={draw=bcArrow,line width=1.6pt,-{Latex[length=7.5pt,width=7pt]}},
  bc title/.style={font=\BCserif{29}\bfseries,text height=22pt,text depth=6pt},
  bc condition/.style={rounded corners=6pt,line width=1pt},
  bc snippet/.style={fill=bcEditor,draw=bcEditorEdge,line width=.8pt,rounded corners=6pt}
]
\path[use as bounding box,fill=white] (-1,-1) rectangle (1521,417);

\foreach \X in {0,514,1028}{
  \begin{scope}[shift={(\X,0)}]
    \path[fill=white,rounded corners=7pt] (0,0) rectangle (492,416);
    \begin{scope}
      \clip[rounded corners=7pt] (0,0) rectangle (492,416);
      \fill[bcHeader] (0,0) rectangle (492,58);
    \end{scope}
    \draw[bcRule,line width=.8pt] (.7,58) -- (491.3,58);
    \draw[bcBorder,line width=1pt,rounded corners=7pt] (0,0) rectangle (492,416);
  \end{scope}
}
\node[bc title] at (246,30) {Python source lines};
\node[bc title] at (760,30) {Same syntax\,\textrightarrow\,one bucket};
\node[bc title] at (1274,30) {Low-entropy bucket\,\textrightarrow\,short code};

\draw[bc flow] (494.8,224) -- (511.2,224);
\draw[bc flow] (1008.8,224) -- (1025.2,224);

\begin{scope}
  \path[fill=bcEditor,rounded corners=8pt] (18,80) rectangle (474,368);
  \begin{scope}
    \clip[rounded corners=8pt] (18,80) rectangle (474,368);
    \fill[bcEditorTop] (18,80) rectangle (474,114);
    \fill[bcEditor] (18,80) rectangle (183,114);
    \fill[bcFunction] (18,80) rectangle (183,82);
    \fill[bcActiveRow] (18,243) rectangle (474,299);
    \fill[bcFunction] (18,257) rectangle (20,285);
  \end{scope}
  \draw[bcEditorEdge,line width=.9pt,rounded corners=8pt] (18,80) rectangle (474,368);
  \draw[bcEditorRule,line width=.65pt] (18.5,114) -- (473.5,114);
  \draw[bcEditorRule,line width=.7pt] (62,131) -- (62,351);

  \draw[bcFunction,line width=1.25pt] (33,91) -- (40,91) -- (44,95) -- (44,105) -- (33,105) -- cycle;
  \draw[bcFunction,line width=1pt] (40,91) -- (40,95) -- (44,95);
  \node[anchor=base west,font=\BCsans{14.7},text=bcCode] at (54,103) {parser.py};
  \draw[bcCodeMuted,line width=1.1pt] (165,95) -- (171,101) (165,101) -- (171,95);
  \foreach \X in {444,450,456}{\fill[bcCodeMuted] (\X,98) circle (1.15pt);}

  \draw[bcEditorRule,line width=.9pt] (70,188) -- (70,347);
  \draw[bcEditorRule,line width=.9pt] (114,247) -- (114,292);
  \foreach \Y in {196,255,314}{
    \path[fill=bcIndent,rounded corners=2pt] (74,\Y) rectangle ++(15,32);
  }
  \path[fill=bcIndent,rounded corners=2pt] (118,255) rectangle (133,287);

  \foreach \N/\Y in {118/160,119/219,120/278,121/337}{
    \node[anchor=base east,font=\BCmono{18.5},text=bcCodeMuted] at (52,\Y) {\N};
  }
  \BCcode{70}{160}{22}{\BCkw{def}\ \BCfn{parse\_rows}\BCpn{(}\BCvar{path}\BCpn{):}}
  \BCcode{114}{219}{22}{\BCkw{if}\ \BCvar{path}\ \BCkw{is}\ \BCconst{None}\BCpn{:}}
  \BCcode{158}{278}{22}{\BCkw{return}\ \BCpn{[]}}
  \BCcode{114}{337}{22}{\BCkw{return}\ \BCfn{read\_rows}\BCpn{(}\BCvar{path}\BCpn{)}}

  \foreach \Y in {212,271}{
    \path[fill=bcPositiveWell,rounded corners=5pt] (416,{\Y-15}) rectangle (460,{\Y+15});
    \node[anchor=base,font=\BCmono{21}\bfseries,text=bcPositive] at (438,{\Y+7}) {+4};
  }
  \path[fill=bcNegativeWell,rounded corners=5pt] (416,315) rectangle (460,345);
  \node[anchor=base,font=\BCmono{21}\bfseries,text=bcNegative] at (438,337) {-4};
\end{scope}

\begin{scope}[shift={(514,0)}]
  \path[bc condition,fill=bcPeach,draw=bcPeachEdge] (17,76) rectangle (159,151);
  \path[bc condition,fill=bcBlue,draw=bcBlueEdge] (175,76) rectangle (317,151);
  \path[bc condition,fill=bcMint,draw=bcMintEdge] (333,76) rectangle (475,151);

  \BClabel{88}{107}{22}{prev ends}
  \BClabel{88}{132}{22}{with {\BCmono{21}\textcolor{bcPeachInk}{\textquotesingle:\textquotesingle}}}
  \BClabel{246}{97}{22}{current}
  \BClabel{246}{120}{22}{starts with}
  \node[anchor=base,font=\BCmono{20},text=bcBlueInk] at (246,143) {return};
  \BClabel{404}{97}{22}{prev indent}
  \BClabel{404}{120}{22}{4--7}
  \BClabel{404}{143}{22}{spaces}

  \foreach \X in {88,246,404}{\draw[bc branch] (\X,156) -- (\X,175);}

  \foreach \X in {17,175,333}{
    \path[bc snippet] (\X,180) rectangle ++(142,88);
    \node[anchor=base west,font=\BCmono{16.8},text=bcCodeMuted] at ({\X+9},198) {...};
    \node[anchor=base west,font=\BCmono{16.8},text=bcCodeMuted] at ({\X+9},260) {...};
  }
  \BCcode{26}{219}{16.8}{\BCkw{for}\ \BCvar{x}\ \BCkw{in}\ \BCvar{items}\BCpn{:}}
  \BCcode{59.6}{240}{16.8}{\BCkw{return}\ \BCvar{x}}
  \BCcode{184}{219}{16.8}{\BCkw{if}\ \BCvar{cond}\BCpn{:}}
  \BCcode{217.6}{240}{16.8}{\BCkw{return}\ \BCpn{[]}}
  \BCcode{342}{219}{16.8}{\BCkw{try}\BCpn{:}}
  \BCcode{375.6}{240}{16.8}{\BCkw{return}\ \BCconst{None}}

  \draw[bc branch] (88,274.5) -- (106,292.5);
  \draw[bc branch] (246,274.5) -- (246,292.5);
  \draw[bc branch] (404,274.5) -- (386,292.5);

  \path[fill=bcLavender,draw=bcLavenderEdge,line width=1.1pt,rounded corners=9pt] (17,299) rectangle (475,382);
  \node[font=\BCserif{24.5}\bfseries] at (246,316) {same syntax bucket};
  \path[fill=white,draw=bcWellEdge,line width=.9pt,rounded corners=7pt] (27,337) rectangle (465,373);
  \foreach \X/\D in {64/+4,116/+4,168/+4,220/+8,272/+4,324/+4,376/+4}{
    \node[font=\BCserif{23}\bfseries,text=bcGreenInk] at (\X,355) {\D};
  }
  \node[font=\BCmono{22},text=bcMuted] at (428,355) {...};
  \node[font=\BCserif{21.5},text=bcMuted] at (246,399) {many similar cases};
\end{scope}

\begin{scope}[shift={(1028,0)}]
  \node[font=\BCserif{27}] at (176.5,101) {mostly +4};

  \def\BCHistMax{840}
  \def\BCCountPlusFour{760}
  \def\BCCountPlusEight{100}
  \def\BCCountZero{50}
  \def\BCCountMinusFour{50}
  \foreach \V in {0,200,400,600,800}{
    \pgfmathsetmacro{\YY}{311-174*(\V/\BCHistMax)}
    \ifnum\V>0
      \draw[bcGrid,line width=.7pt] (79,\YY) -- (274,\YY);
    \fi
    \draw[bcMuted,line width=.8pt] (75,\YY) -- (79,\YY);
    \node[anchor=east,font=\BCserif{19},text=bcMuted] at (68,\YY) {\V};
  }
  \node[rotate=90,font=\BCserif{21},text=bcMuted] at (27,224) {count};

  \foreach \X/\V/\Fill/\Stroke in {
    104.5/\BCCountPlusFour/bcBarStrong/bcBarStrongEdge,
    152.5/\BCCountPlusEight/bcBarSoft/bcBarSoftEdge,
    200.5/\BCCountZero/bcBarSoft/bcBarSoftEdge,
    248.5/\BCCountMinusFour/bcBarRed/bcBarRedEdge}{
    \pgfmathsetmacro{\YY}{311-174*(\V/\BCHistMax)}
    \path[fill=\Fill,draw=\Stroke,line width=.85pt] ({\X-16},\YY) rectangle ({\X+16},311);
  }
  \draw[bcInk,line width=1.25pt] (79,137) -- (79,311) -- (274,311);
  \foreach \X/\V in {104.5/+4,152.5/+8,200.5/0,248.5/\textminus4}{
    \draw[bcMuted,line width=.8pt] (\X,311) -- (\X,316);
    \node[font=\BCserif{21}] at (\X,336) {\V};
  }
  \node[font=\BCserif{21.5},text=bcMuted] at (176.5,367) {indentation delta};

  \draw[bc flow,line width=2pt] (286.5,224) -- (318.5,224);
  \path[fill=bcEncoder,draw=bcEncoderEdge,line width=1.1pt,rounded corners=10pt] (331,164) rectangle (477,284);
  \node[font=\BCserif{29}] at (404,194) {rANS};
  \path[fill=bcBitWell,draw=bcEncoderEdge,line width=.85pt,rounded corners=6pt] (343,219) rectangle (465,268);
  \node[font=\BCmono{22},text=bcMuted] at (404,243.5) {010110...};
  \node[font=\BCserif{19.3},text=bcMuted] at (404,305) {(compact bitstream)};
\end{scope}
\end{tikzpicture}%
\endgroup%
}%
  \endgroup

  \caption{Syntax-conditioned indentation coding for Python. The histogram is schematic.}
  \label{fig:python-indentation}
\end{figure}

The decoder can follow the same logic. Once it has reconstructed the line contents, it has the clues needed to restore the leading spaces, as illustrated by the \texttt{return} example in Fig.~\ref{fig:python-indentation}. There is therefore no need to store a context identifier for each line. Astra combines this indentation scheme with dictionary substitution and LZ parsing, producing a codec that achieves an 8\% higher compression factor than XZ-9 while decoding 76\% faster (Fig.~\ref{fig:block-comparison}(b)).

\subsection{Database String Columns}
\label{sec:results-dbtext}

DBText, unlike the other datasets, gives Astra a varied collection of data to work with: 23 columns ranging from multilingual text and literary works to identifiers and coordinates, and each run produces one encoder and decoder for all of them. One might expect the resulting codec to use a separate compression algorithm for every column, but in \Aone{}, Astra reuses the same phrase encoder and decoder across nine columns, handling Hamlet and Faust alongside Chinese and Japanese text and database comments. Each column still has its own learned dictionary, so the shared algorithm can adapt to the phrases that recur in that input. At the same time, Astra keeps the overhead down by removing unused phrases or grammar rules. On top of this, \Afour{} merges duplicate entries within each column's dictionary and compresses the dictionary itself.

\begin{figure}[!htbp]
  \centering
  \begingroup
  \resizebox{\linewidth}{!}{\begingroup%
\begin{tikzpicture}[x=0.333333pt,y=-0.333333pt,line cap=round,line join=round]
\providecommand{\UUIDPanelGap}{32}
\providecommand{\UUIDCorrectBits}{1}
\providecommand{\UUIDShowTitle}{0}

\definecolor{UUIDInk}{HTML}{061630}
\definecolor{UUIDText}{HTML}{40516A}
\definecolor{UUIDOrange}{HTML}{FF4808}
\definecolor{UUIDBlue}{HTML}{0876FF}
\definecolor{UUIDGreen}{HTML}{008E0D}
\definecolor{UUIDExampleBorder}{HTML}{BECCDD}
\definecolor{UUIDExampleHeader}{HTML}{EBEFF3}
\definecolor{UUIDOrangeBorder}{HTML}{FFD5BF}
\definecolor{UUIDOrangeBody}{HTML}{FFF7F2}
\definecolor{UUIDOrangeHeader}{HTML}{FFE4D5}
\definecolor{UUIDBlueBorder}{HTML}{BDDEFF}
\definecolor{UUIDBlueBody}{HTML}{F0F8FF}
\definecolor{UUIDBlueHeader}{HTML}{D8ECFD}
\definecolor{UUIDGreenBorder}{HTML}{B7F4BC}
\definecolor{UUIDGreenBody}{HTML}{F3FEF5}
\definecolor{UUIDGreenHeader}{HTML}{D9FBDD}
\definecolor{UUIDHighlightTop}{HTML}{B1F7B0}
\definecolor{UUIDHighlightBottom}{HTML}{A9F3A8}
\definecolor{UUIDDivider}{HTML}{C6D4DB}
\definecolor{UUIDArrowTop}{HTML}{98A9BD}
\definecolor{UUIDArrowBottom}{HTML}{7B8BA1}

\def\UUIDExampleX{14}
\def\UUIDExampleWidth{682}
\def\UUIDArrowGap{84}
\def\UUIDPrefixWidth{414}
\def\UUIDConstantWidth{252}
\def\UUIDBitsWidth{706}
\def\UUIDTop{184}
\def\UUIDHeight{264}
\def\UUIDHeaderHeight{72}
\pgfmathsetmacro{\UUIDBottom}{\UUIDTop+\UUIDHeight}
\pgfmathsetmacro{\UUIDBodyTop}{\UUIDTop+\UUIDHeaderHeight}
\pgfmathsetmacro{\UUIDBodyCentre}{(\UUIDBodyTop+\UUIDBottom)/2}
\pgfmathsetmacro{\UUIDPrefixX}{\UUIDExampleX+\UUIDExampleWidth+\UUIDArrowGap}
\pgfmathsetmacro{\UUIDConstantX}{\UUIDPrefixX+\UUIDPrefixWidth+\UUIDPanelGap}
\pgfmathsetmacro{\UUIDBitsX}{\UUIDConstantX+\UUIDConstantWidth+\UUIDPanelGap}
\pgfmathsetmacro{\UUIDFigureWidth}{\UUIDBitsX+\UUIDBitsWidth+14}
\pgfmathsetmacro{\UUIDArrowX}{\UUIDExampleX+\UUIDExampleWidth+\UUIDArrowGap/2}
\ifnum\UUIDShowTitle=1\relax
  \def\UUIDCropTop{0}
\else
  \def\UUIDCropTop{174}
\fi

\tikzset{
  uuid/base/.style={inner sep=0pt,outer sep=0pt,text=UUIDText,
    font=\fontfamily{qtm}\selectfont},
  uuid/header/.style={uuid/base,
    font=\fontsize{12.4pt}{14pt}\fontfamily{qtm}\bfseries\selectfont,
    text height=8.6pt,text depth=2.4pt},
  uuid/data/.style={uuid/base,
    font=\fontsize{12pt}{14pt}\fontfamily{qtm}\selectfont,
    text height=8.1pt,text depth=0pt},
  uuid/prefix/.style={uuid/base,
    font=\fontsize{9.6pt}{12pt}\fontfamily{qtm}\selectfont,
    text height=6.6pt,text depth=1.7pt},
  uuid/formula/.style={uuid/base,
    font=\fontsize{9.7pt}{12pt}\fontfamily{qtm}\selectfont},
  uuid/bit/.style={uuid/base,anchor=base,
    font=\fontsize{10.9pt}{12pt}\fontfamily{qtm}\selectfont},
  uuid/key/.style={uuid/base,anchor=base east,
    font=\fontsize{9.6pt}{11.5pt}\fontfamily{qtm}\bfseries\selectfont},
  uuid/value/.style={uuid/base,anchor=base west,
    font=\fontsize{9.3pt}{11.5pt}\fontfamily{qtm}\selectfont}
}

\newcommand{\UUIDPanel}[7]{%
  \begin{scope}
    \clip[rounded corners=4.5pt]
      (#1,\UUIDTop) rectangle ({#1+#2},\UUIDBottom);
    \fill[#4] (#1,\UUIDTop) rectangle ({#1+#2},\UUIDBottom);
    \fill[#5] (#1,\UUIDTop) rectangle ({#1+#2},\UUIDBodyTop);
  \end{scope}
  \draw[draw=#3,line width=0.65pt,rounded corners=4.5pt]
    (#1,\UUIDTop) rectangle ({#1+#2},\UUIDBottom);
  \node[uuid/header,text=#6]
    at ({#1+#2/2},{\UUIDTop+\UUIDHeaderHeight/2}) {#7};
}

\newcommand{\UUIDValue}[2]{%
  {\color{UUIDOrange}#1}%
  {\color{UUIDText}-}%
  {\color{UUIDBlue}abcd-1234}%
  {\color{UUIDText}-}%
  {\color{UUIDGreen}#2}%
  {\color{UUIDText}-001122334455}%
}

  \path[use as bounding box]
    (0,\UUIDCropTop) rectangle (\UUIDFigureWidth,462);

  \ifnum\UUIDShowTitle=1\relax
    \node[uuid/base,anchor=base,text=UUIDInk,
      font=\fontsize{19.6pt}{23pt}\fontfamily{qtm}\bfseries\selectfont]
      at ({\UUIDFigureWidth/2},67)
      {UUID column\, $\boldsymbol{\rightarrow}$\, compact representation};
    \node[uuid/base,anchor=base,
      font=\fontsize{14pt}{17pt}\fontfamily{qtm}\bfseries\selectfont]
      at ({\UUIDFigureWidth/2},126)
      {Discovered structure across rows};
  \fi

  \UUIDPanel{\UUIDExampleX}{\UUIDExampleWidth}
    {UUIDExampleBorder}{white}{UUIDExampleHeader}{UUIDInk}{Example UUID values}
  \UUIDPanel{\UUIDPrefixX}{\UUIDPrefixWidth}
    {UUIDOrangeBorder}{UUIDOrangeBody}{UUIDOrangeHeader}{UUIDOrange}{Arithmetic prefix}
  \UUIDPanel{\UUIDConstantX}{\UUIDConstantWidth}
    {UUIDBlueBorder}{UUIDBlueBody}{UUIDBlueHeader}{UUIDBlue}{Constant field}
  \UUIDPanel{\UUIDBitsX}{\UUIDBitsWidth}
    {UUIDGreenBorder}{UUIDGreenBody}{UUIDGreenHeader}{UUIDGreen}{Varying bits}

  \node[uuid/data] at ({\UUIDExampleX+\UUIDExampleWidth/2},{\UUIDBodyCentre-48})
    {\UUIDValue{10000000}{8a20}};
  \node[uuid/data] at ({\UUIDExampleX+\UUIDExampleWidth/2},\UUIDBodyCentre)
    {\UUIDValue{10000010}{8a24}};
  \node[uuid/data] at ({\UUIDExampleX+\UUIDExampleWidth/2},{\UUIDBodyCentre+48})
    {\UUIDValue{10000020}{8a28}};

  \shade[top color=UUIDArrowTop,bottom color=UUIDArrowBottom]
    ({\UUIDArrowX-24},{\UUIDBodyCentre-11}) --
    ({\UUIDArrowX+1},{\UUIDBodyCentre-11}) --
    ({\UUIDArrowX+1},{\UUIDBodyCentre-28}) --
    ({\UUIDArrowX+27},\UUIDBodyCentre) --
    ({\UUIDArrowX+1},{\UUIDBodyCentre+28}) --
    ({\UUIDArrowX+1},{\UUIDBodyCentre+11}) --
    ({\UUIDArrowX-24},{\UUIDBodyCentre+11}) -- cycle;

  \node[uuid/prefix,text=UUIDOrange]
    at ({\UUIDPrefixX+\UUIDPrefixWidth/2},{\UUIDBodyCentre-25})
    {10000000,\enspace 10000010,\enspace 10000020};
  \ifnum\UUIDCorrectBits=1\relax
    \node[uuid/formula]
      at ({\UUIDPrefixX+\UUIDPrefixWidth/2},{\UUIDBodyCentre+27})
      {$\mathrm{0x10000000}+\mathrm{0x10}\times[0,\,1,\,2]$};
  \else
    \node[uuid/formula]
      at ({\UUIDPrefixX+\UUIDPrefixWidth/2},{\UUIDBodyCentre+27})
      {$10000000+10\times[0,\,1,\,2]$};
  \fi
  \node[uuid/data,text=UUIDBlue]
    at ({\UUIDConstantX+\UUIDConstantWidth/2},\UUIDBodyCentre)
    {abcd-1234};

  \def\UUIDBitPitch{16.5}
  \ifnum\UUIDCorrectBits=1\relax
    \def\UUIDBitCount{16}
  \else
    \def\UUIDBitCount{17}
  \fi
  \pgfmathsetmacro{\UUIDBitFirstX}{\UUIDBitsX+172-(\UUIDBitCount-1)*\UUIDBitPitch/2}
  \pgfmathsetmacro{\UUIDHighlightLeft}{\UUIDBitFirstX+12*\UUIDBitPitch-\UUIDBitPitch/2-4}
  \ifnum\UUIDCorrectBits=1\relax
    \pgfmathsetmacro{\UUIDHighlightRight}{\UUIDHighlightLeft+2*\UUIDBitPitch+8}
  \else
    \pgfmathsetmacro{\UUIDHighlightRight}{\UUIDHighlightLeft+\UUIDBitPitch+8}
  \fi
  \shade[rounded corners=2.5pt,
      top color=UUIDHighlightTop,bottom color=UUIDHighlightBottom]
    (\UUIDHighlightLeft,{\UUIDBodyTop+18})
      rectangle (\UUIDHighlightRight,{\UUIDBottom-18});

  \newcommand{\UUIDBitRow}[2]{%
    \foreach \digit [count=\digitindex from 0] in {#2} {%
      \node[uuid/bit]
        at ({\UUIDBitFirstX+\digitindex*\UUIDBitPitch},{#1+11.2}) {\digit};
    }%
  }
  \ifnum\UUIDCorrectBits=1\relax
    \UUIDBitRow{\UUIDBodyCentre-48}{1,0,0,0,1,0,1,0,0,0,1,0,0,0,0,0}
    \UUIDBitRow{\UUIDBodyCentre}{1,0,0,0,1,0,1,0,0,0,1,0,0,1,0,0}
    \UUIDBitRow{\UUIDBodyCentre+48}{1,0,0,0,1,0,1,0,0,0,1,0,1,0,0,0}
  \else
    \UUIDBitRow{\UUIDBodyCentre-48}{1,0,0,0,1,0,1,0,0,0,0,0,0,0,0,0,0}
    \UUIDBitRow{\UUIDBodyCentre}{1,0,0,0,1,0,1,0,0,0,1,0,1,0,1,0,0}
    \UUIDBitRow{\UUIDBodyCentre+48}{1,0,0,0,1,0,1,0,0,0,1,0,1,1,0,0,0}
  \fi

  \draw[UUIDDivider,line width=0.65pt]
    ({\UUIDBitsX+350},{\UUIDBodyTop+36}) --
    ({\UUIDBitsX+350},{\UUIDBottom-36});
  \node[uuid/key] at ({\UUIDBitsX+451},{\UUIDBodyCentre-24+10}) {fixed:};
  \node[uuid/value] at ({\UUIDBitsX+462},{\UUIDBodyCentre-24+10})
    {100010100010-{}-00};
  \node[uuid/key] at ({\UUIDBitsX+451},{\UUIDBodyCentre+24+10}) {mask:};
  \node[uuid/value] at ({\UUIDBitsX+462},{\UUIDBodyCentre+24+10})
    {0000000000001100};
\end{tikzpicture}%
\endgroup%
}%
  \endgroup

  \caption{Illustrative UUID structure: an arithmetic prefix, a repeated field, and varying bits.}
  \label{fig:dbtext-uuid}
\end{figure}
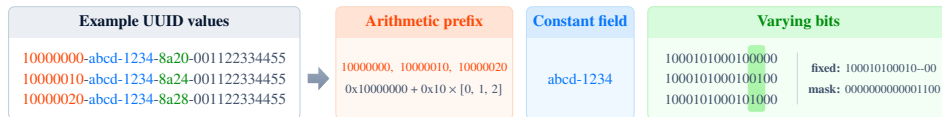

While Astra reuses the same phrase encoder and decoder across several text columns, it finds a different opportunity in the UUID column of DBText (Fig.~\ref{fig:dbtext-uuid}): identifiers that look different from row to row still share parts that need to be stored only once. Their middle fields and some suffix bits stay constant across the column, so these become shared metadata. In \Atwo{}, Astra also notices a pattern in the leading field that can be described with a shared base and step. With these patterns captured, each row needs only a count of steps above the base and the suffix bits that vary. Packing the suffix bit by bit allows even a changing hexadecimal character to occupy fewer than four bits, and the fixed bit width of the resulting records lets the decoder locate a requested row directly.

\section{Related Work}\label{sec:related-work}

Within a database, saving space is only part of a compressor's job. The compressed data must also support fast decoding and access to individual values. FSST addresses these requirements with a shared static symbol table, while allowing each string to be decoded independently~\cite{fsst2020}. OnPair builds its dictionary by merging frequent adjacent substrings, capturing recurring patterns in the input~\cite{onpair2025}. OnPair+ builds on this design with optimizations that improve compression factor and decompression throughput~\cite{tokenTables2026}.

Language models offer another way to capture patterns in text. LLMZip and FineZip use model predictions to encode text compactly~\cite{llmzip2023,finezip2024}. However, reconstruction also requires model inference, so the storage savings come with additional computational costs~\cite{waitingDecompress2026}. CandleZip brings agents into this process through hints that guide compression and accounts for the information needed to replay them during decoding~\cite{agenticCompression2025}.

An agent can also help develop the compression algorithm itself. The MinEvolve prototype starts with an LZ77-style compressor and uses a language model to propose changes to its code~\cite{minevolve2025}. Candidates are evaluated by compression ratio, and those that fail reconstruction checks are rejected. Our study follows this direction with agents that develop complete native codecs for fixed datasets, pursuing different objectives for compression factor and decompression speed. Once a codec has been constructed and validated, encoding and decoding no longer require the model.

\section{Conclusions \& Future Work}\label{sec:conclusion}

We have shown that frontier models can synthesize compression codecs that extend the compression factor--decompression speed Pareto front across five workloads. The synthesized codecs are guaranteed to be correct, since reconstruction of a fixed input can be checked byte for byte, allowing us to reject erroneous attempts. Once validated, these codecs also run independently of the model.

At the same time, our ablations show that the synthesis process matters, with the model, performance targets, and available tools influencing what agents discover. Our results thus establish what current agents can achieve under this protocol, while leaving open how much further they could go. Future work could therefore explore this headroom through systematic prompt optimization, controlled ablations, and richer tools for data inspection, profiling, and experimental feedback. Beyond compression, we could also look for DBMS components whose results can be exhaustively verified within a clearly defined scope, giving agents the same freedom to specialize while preserving correctness.

\printbibliography

\end{document}